\documentstyle[prl,aps]{revtex}
\begin{document}
\author{T. Giamarchi}
\address{Laboratoire de Physique des Solides, Universit{\'e} Paris-Sud,\\
B{\^a}t. 510, 91405 Orsay, France\cite{junk}}
\author{P. Le Doussal}
\address{CNRS-Laboratoire de Physique Th\'eorique de l'Ecole\\
Normale Sup\'erieure, 24 rue Lhomond, F-75231 Paris\cite{frad}}
\title{Moving glass phase of driven lattices}
\date{\today}
\maketitle

\begin{abstract}
We study periodic lattices, such as vortex lattices, driven by an external
force in a random pinning potential. We show that effects of static disorder
persist even at large velocity. It results in a novel
moving glass state which has analogies with the static Bragg
glass. The lattice flows through well-defined, elastically coupled, {\it %
static} channels. We predict barriers to transverse motion resulting in
finite transverse critical current. Experimental tests of the theory
are proposed.
\end{abstract}

\pacs{to be added}

\narrowtext

An open question is to
understand the effect of static substrate disorder on periodic 
media such as vortex lattices \cite{blatter_vortex_review},
charge density waves (CDW) \cite{gruner_revue_cdw},
Wigner crystals \cite{wigner_andrei}, colloids \cite{murray_colloid_prb},
magnetic bubbles \cite{seshadri_bubbles_long}.
Numerous experiments study such elastic systems in motion under
an applied force produced by 
a current (vortex lattices), a voltage (CDW),
an electric field (colloids) and 
a magnetic field gradient (magnetic bubbles).
It is therefore important to describe the physical properties
of both the static and moving lattices.
The statics of vortex
lattices has been much investigated recently and it is generally agreed 
that disorder leads to a glass state with diverging
barriers, pinning and loss of translational order.
The precise nature of the
glass state however has been the subject of much debate
\cite{feigelman_collective,fisher_vortexglass_long,nattermann_vortex}
in particular concerning the decay of translational order
and the presence of topological defects. It was
shown recently \cite{giamarchi_vortex_global} within an elastic theory
that, because of the periodicity of the lattice, the decay of translational
order is only algebraic and that the resulting glass phase still 
exhibits divergent Bragg peaks. We argued that
at weak disorder, a Bragg glass exists without equilibrium 
dislocations and also that such a glass will
undergo a transition into a strongly disordered 
vortex glass containing topological defects, or a pinned liquid, 
upon increase of disorder or field \cite{giamarchi_vortex_global}.
This is compatible with recent decoration and neutron
experiments \cite{yaron_neutron,giamarchi_comment-neutrons}
and with the behaviour of the critical point \cite{safar_tricritical_prl}
in the phase diagram of vortex lattices where a transition between
two different glass states is observed upon raising the field
\cite{khaykovich_zeldov}.

It is thus crucial to determine
how much of the glassy properties of the static system
remain once the lattice is set in motion, and how translational
and topological order behave.
At large velocity $v$ it was expected that since the pinning force 
on a given vortex varies rapidly, disorder would
produce little effect. Perturbation theories in 
disorder and $1/v$ were thus developped
\cite{hauger_schmidt,larkin_ovchinnikov}
to compute velocity as a function of the external force,
and to estimate critical currents.
Recently, Koshelev and Vinokur
\cite{koshelev_dynamic} have extended the
perturbation theory of \cite{hauger_schmidt,larkin_ovchinnikov}
to compute vortex displacements $u$.
They concluded that at low $T$ and above a certain velocity 
the moving lattice is a crystal at an effective temperature 
$T^{\prime }=T+T_{sh}$. 
Several experiments indeed suggest that a fast 
moving lattice is more ordered \cite{thorel_vortex,yaron_neutron}.
The effect of pinning
can be described \cite{koshelev_dynamic} 
by some effective shaking temperature
$T_{sh}\sim 1/v^2$ defined by the relation
$\langle |u(q)|^2 \rangle = T_{sh}/c_{66} q^2$
This would suggest bounded displacements in $d>2$ and
absence of glassy properties in the moving solid.

In this Letter we reconsider this problem. We show that in the case
of a moving lattice the perturbation theory of \cite{koshelev_dynamic} 
breaks down, even at large $v$.
The physical reason is that some modes
of the disorder are not affected by the
motion and {\it static} disorder is still present in the moving system.
As a result the moving lattice is in fact a {\it moving glass}.
Since translational order in the moving frame decays 
and relative displacements are not bounded, such a
phase cannot be described by simple perturbation theory \cite{perturb}.
As in the statics, {\it periodicity} is crucial and the moving
lattice has a completely different behaviour than other
driven systems such as manifolds.
The physics of this new phase can be described
in terms of {\it elastic channels}.
When the force is applied along a principal lattice direction 
the rows of the lattice
flow along well-defined, nearly parallel, preferred paths in the
pinning potential. 
The manifold of these optimal channels
(lines for 2D lattice and sheets for 3D vortex line lattice) which
exhibits a roughness that we estimate, are a purely static and
reproducible feature of the disorder configuration.
We also predict that the moving glass exhibits
barriers to an additional small transverse force and compute
the associated transverse critical current. The other modes of the
disorder are suppressed by motion and give
rise to an additional wiggling motion of the particles 
around the static channel configuration, which can be treated
in perturbation.

We now derive the equation of motion for a lattice submitted
to external force $F$. We denote by $R_i(t)$ the position of an individual vortex
in the laboratory frame. The lattice as a whole moves
with a velocity $v$. We thus introduce the displacements $R_i(t)=R_i^0+vt+u_i(t)$
where the $R_i^0$ denote the equilibrium positions in
the perfect lattice with no disorder. $u_i$ represent the displacements in
the moving frame. We consider in the following the elastic limit
in the absence of topological defects, thus assuming 
$|u_i - u_{i+1}| \ll a$ where $a$ is the lattice spacing, an 
assumption which may be checked self-consistently. One then
takes the continuum limit $u_i(t)\to u(r,t)$,
where $u(r,t)$ is a smoothly varying $n$-component vector field,
which components we denote by $u_\alpha(r,t)$.
It is convenient here to express the displacement field
$u_\alpha(r,t)$ in terms of the coordinates $(r,t)$ of the laboratory frame.
The equation of motion in the laboratory frame is then:
\begin{equation}
\eta \partial_t u_\alpha +\eta v \cdot \nabla u_\alpha = c\nabla^2 u_\alpha 
+ F^{\text{pin}}_{\alpha} 
+ F_{\alpha}-\eta v_{\alpha} + \zeta_{\alpha}
\label{eqmotion}
\end{equation}
where $\eta$ is the friction coefficient, 
$F^{\text{pin}}_{\alpha}(r,t) = - \delta {\cal E}/\delta u_{\alpha}(r,t)$
is the pinning force, ${\cal E}[u(r,t)]$ is the pinning energy, and the thermal noise
satisfies $\overline{\zeta_{\alpha} (r,t)\zeta_{\beta} (r^{\prime},t^{\prime})} 
= 2 T \delta_{\alpha \beta}
\delta^d(r-r^{\prime})\delta(t-t^{\prime})$. For clarity we
use here an isotropic elastic constant $c$.
The realistic case, discussed at the end, has the same large distance
physics. The term $\eta v \cdot \nabla u_\alpha$ comes from expressing the 
displacement field in the laboratory
frame and $-\eta v_{\alpha}$ is the average friction. 
$v$ is determined by the condition that the average of $u$ is zero.
(1) is exact up to higher powers of derivatives of $u$, negligible
in the elastic limit.
The pinning energy can be expressed in terms of the vortex density $%
\rho(r,t)=\sum_i\delta (r-R_i^0-vt-u_i(t))$. One has ${\cal E}[u(r,t)]=\int
d^dr\rho(r,t)V(r)$ where the random potential has correlations $\langle
V(r)V(r^{\prime})\rangle = \Delta(r-r^{\prime})$ 
of range $r_f$. Since even for smooth displacement fields
the density is a series of delta peaks, the continuum limit for
${\cal E}[u]$ should be performed by distinguishing \cite{giamarchi_vortex_global}
the various Fourier components of the density
$\rho (r,t)=\rho _0(1-\nabla \cdot u + \sum_{K\neq 0} \exp(iK \cdot
(r-vt-u(r,t)))$ where $K$ spans the reciprocal lattice and $\rho_0$ is the 
average density. Using this decomposition 
in (\ref{eqmotion}) the force due to disorder naturally splits
into a {\it static} and a time-dependent part: \widetext
\begin{eqnarray}
&&\eta \partial_t u_\alpha + \eta v \cdot \nabla u_\alpha = c\nabla^2 u_\alpha
+F_\alpha^{\text{stat}}(r,u(r,t)) + F_\alpha^{\text{dyn}}(r,t,u(r,t)) + F_\alpha -
\eta v_\alpha + \zeta_\alpha(r,t)  \label{eqmotion2} \\
&& F_\alpha ^{\text{stat}}(r,u) = V(r)\rho_0\sum_{K.v=0} iK_\alpha 
e^{iK\cdot(r-u)}  - \rho_0 \nabla_\alpha V(r) , \qquad
F_\alpha^{\text{dyn}}(r,t,u) = V(r)\rho_0\sum_{K.v\neq 0} iK_\alpha 
e^{iK\cdot(r-vt-u)} \nonumber
\end{eqnarray}
\narrowtext
The static part of the random force comes from the modes
such that $K.v=0$ which exist for any 
direction of the velocity commensurate with the lattice.
The maximum effect is obtained for $v$ parallel to one
principal lattice direction, the situation we study now.
Reflection symmetry then imposes that $v$ and $F$ 
are aligned along direction $x$,
the $d-1$ transverse directions being denoted by $y$.
$F_\alpha ^{\text{stat}}$ gives the dominant contribution 
to the lattice deformations. 
In first approximation we drop
$F_\alpha ^{\text{dyn}}$ and
solve the remaining static problem (leading to a reference
ground state at $T=0$). $F_\alpha^{\text{dyn}}$ gives
additional fluctuations 
on top of this ground state, estimated below.
The static term $\rho_0 \nabla_\alpha V$, which comes from the Fourier 
components $k \ll 1/a$ of the disorder, produces alone only bounded displacements 
for $d > 1$. Thus, as for the 
non-moving lattice \cite{giamarchi_vortex_global} for $d>2$, it
does not change the large scale physics and we drop it.
Since $F_\alpha^{\text{stat}}$ is now along $y$, and depends only
on $u_y$, $u_x=0$ in the ground state.

The most important terms in (1) thus lead to
the following equation of motion in the laboratory frame which involves
only the {\it transverse} displacements $u_y$:
\begin{eqnarray}
\eta \partial_t u_y +\eta v\partial_x u_y &=& c\nabla^2 u_y
+ F^{\text{stat}}(r,u_y(r,t)) + \zeta_y (r,t) \nonumber \\
F^{\text{stat}}(x,y,u_y) &=& V(x,y) \rho_0 \sum_{Ky \ne 0} K_y
\sin{K_y(u_y-y)}
\label{staticequ}
\end{eqnarray}
This is now a non-trivial static disordered model and one expects 
a glass phase at low temperature, 
with pinning of the field $u(r,t)$ into preferred configurations. 
Thus, the moving vortex configurations 
can be described in terms of {\it static channels}
that are the easiest paths where particle follow
each other in their motion. 
Channels in the elastic flow regime behave differently
than the one introduced to describe slow plastic motion between 
pinned islands \cite{jensen}.
In the topologically ordered moving glass
they form a manifold
of elastically coupled, almost parallel lines or sheets (for
vortex lines in $d=3$) directed along $x$ 
and characterized by transverse wandering $u_y$.
In the laboratory frame they are determined by the static disorder 
and do not fluctuate with time. 
In the moving frame, since each
particle is tied to a given channel which is now
moving, it indeed wiggles and dissipates 
but remains highly correlated with the neighbors.
To obtain the roughness of the manifold
of channels we compute the correlator of relative displacements
$B(x,y)=\langle [u(x,y)-u(0,0)]^2\rangle$. A detailed analysis will be presented 
elsewhere \cite{giamarchi_moving_long}. One defines two characteristic lengths
for decay of translational order,
$R_x^a$ and $R_y^a$ along the longitudinal and transverse direction 
by $B(R) \sim a^2$. One expects three regimes.

{\it short scale regime}:
At very short scales one can expand the pinning force
to lowest order in $u$. This gives a simple model where pinning is
described by a random force $F^{\text{stat}}(x)$ independent of $u$
whose correlator is $\langle F^{\text{stat}}(r)F^{\text{stat}}(r')\rangle
=\Delta \delta^d(r-r^{\prime })$ with 
$\Delta = \rho_0^2 \sum_{K_y} K_y^2\Delta _K$. 
This is the dynamic equivalent of the Larkin random
force model and $B = B_{rf} + \langle u^2\rangle_{th}$:
\begin{equation} \label{larkinmove}
B_{rf}(x,y)=\int \frac{dq_x
d^{d-1} q_y}{(2\pi)^d} \frac{\Delta (1-\cos(q_x x+q_y y))}{(\eta v q_x)^2 +
c^2(q_x^2+q_y^2)^2}
\end{equation}
and $\left\langle u^2 \right\rangle_{th}$ is the
thermal displacement. One finds for $x > c/\eta v$
$B(x,y) \sim \Delta \frac{y^{3-d}}{c \eta v}H(c x/\eta v y^2)$
where $H(0)=\text{cst}$ and $H(z)\sim z^{(3-d)/2}$ at large $z$. $x$
scales as $y^2$ and the displacements are very anisotropic.
For $x < c/\eta v$ one has the usual isotropic result.
If one could extrapolate this behaviour
to larger scales it would result in an algebraic decay of 
translational order in $d=3$ ($B(x,y) \sim \log|y|$)
and exponential decay in $d=2$. However,
since the Larkin model rests on the expansion in powers of $u$,
it is valid only as long as $K_{\text{max}} u \ll 1$,
where $K_{\text{max}} \sim 1/r_f$
is the highest Fourier component of the random potential
\cite{giamarchi_vortex_global}.
This defines two lengths
$R_x^c$ and $R_y^c$ such that
$B(R) \sim r_f^2$, below which the Larkin model is valid. At large velocity:
$R_y^c = (r_f^2 v c/\Delta)^{1/(3-d)}$, $R_x^c = v (R_y^c)^2/c$. 
For smaller velocities $v < v^* \sim c(\Delta/c^2 r_f^2)^{1/(4-d)}$
the elastic term dominates and
$R_x^c \sim R_y^c \sim R_{iso}^c$ where $R_{iso}^c$ is the
static pinning length. These
lengths are renormalized by temperature and by 
the dynamical part $u^d$. Note that
this Larkin random force corresponds
formally to the so-called ``random mobility term'' considered
in \cite{krug,balents_dynamics_vortex} and 
by keeping only this term one
misses all the physics of the moving glass, e.g. the channels and 
the transverse barriers. As for the static case 
\cite{giamarchi_vortex_global} the
pinning force in (\ref{staticequ}) 
should be treated to all orders in $u$. Above this length scale
pinning and metastability appear.

{\it intermediate regime}: At intermediate scales $R_y^c < y < R_y^a$
and $R_x^c < x < R_x^a$, the analogous of a random manifold regime 
\cite{feigelman_collective,giamarchi_vortex_global} exists
for which $u_y \sim y^\zeta$. The channels
are determined by optimization of elasticity ($c q^2$ term),
dissipation ($i \eta v q_x$ term) and the random potential seen
independently by each channel in its vicinity. One expects
many metastable nearly optimal configurations in that regime
and glassy behaviour. Flory type arguments suggest that
the scaling properties of this glass are related to the static Bragg
glass by $d\to d+1$ and $n\to n-1$. The former comes from
assuming $q_x\sim q_y^2$ and the latter from $u\to u_y$. 
The Flory estimate is then 
$\zeta^F=\frac{3-d}{n+3}$.

{\it asymptotic regime}:
At large distances $x> R_x^a, y> R_y^a$, in $d=3$ the displacements
have a slower, logarithmic growth.
Estimates a la Fukuyama-Lee then give:
\begin{equation} \label{fukulee2}
R_y^a \sim (a^2 v c/\Delta)^{1/(3-d)}, \qquad
R_x^a = v (R_y^a)^2/c
\end{equation}
The moving glass is highly anisotropic since $R_x^a/R_y^a$ 
diverges as $v\to \infty $.
Its upper critical dimension is $d=3$, 
instead of $d=4$ for the static one. For $d > 3$ the moving system 
is not a glass but a perfect crystal at weak 
disorder or large $v$. For $d \leq 3$ weak disorder 
destroys long range order and results in a moving glass.

As an important consequence of the existence of the moving glass,
barriers for transverse motion exist 
once the pattern of channels is established. Thus the
response to an additional small transverse force $F_y$
is very non linear with activated behaviour. At $T=0$, neglecting the dynamic part of the disorder 
a true transverse critical current $J_y^c$ should exist.
This can be seen by adding a transverse force in (\ref{staticequ}).
For $\eta v < \eta v^* \sim (R^c_{iso}/r_f) F_c$ where $F_c$
is the isotropic critical force, the Larkin domains remain
isotropic
and one expects $J_y^c$ to decrease slowly from $J^c_{iso}$ 
to a fraction of the longitudinal
critical current $J_{iso}^c$ (since only the $K_y$
modes contribute). For $\eta v > \eta v^*$ one expect a
much faster decay and
a naive estimate for $J_y^c$ is obtained by balancing the pinning force
with the transverse force acting on a Larkin domain:
\begin{displaymath}
J_y^c=\frac{c_l}{\phi_0 r_f}\Delta^{1/2} (R_y^c)^{-(d-1)/2}(R_x^c)^{-1/2} \sim
 \widetilde{\Delta }
^{2/(3-d)}
\end{displaymath}
where $\tilde{\Delta }=\Delta /v$ is 
an effective velocity-dependent disorder
and $c_l$ is the speed of light. In $d=3$ it
yields exponential decay with $v$. In practice,
$\eta v^*$ can correspond to a large
driving compared to $F_c$.
The above regimes correspond
to collective  pinning with $R_y^c >a$.
For $R_y^c <a$, i.e 
$v < v_0 = \Delta a^{3-d}/c r_f^2$, single channel pinning
leads to different estimates for $J_y^c$.
Since motion is not modified 
below $J_y^c$, $dv_x/df_y$ also vanishes below the transverse 
critical force. One can expect non linear
effects in the flow along $x$ since channel configuration
is modified when $F_x$ is increased.
A simpler example of transverse barriers
is a lattice driven in a commensurate
washboard potential $V(x,y)=U_0 \cos(K_0 y)- F_x x$.
There it is clear that even in the
moving frame the problem is static
and that the transverse critical force is
$F_y^c \sim U_0 K_0$. Even a 
single particule in the 2D potential
$V(x,y) = f_c \cos(x) + f_c \cos(y) - F_x x$, has finite $F_y^c$.
Its velocity can be computed as in
\cite{ledoussal_vinokur} and becomes for $T \to 0$,
$V_y = (F_y^2 - f_c^2)^{1/2} \theta(F_y-f_c)$,
independent of $F_x$.

To estimate the effects of the time-dependent pinning
force in (\ref{eqmotion2}), we split $u=u^s+u^d$ into a
static $u^s$ and a dynamics $u^d$ part. 
A reasonable estimate is:
\begin{equation}  \label{dynamicequ}
\langle u^d . u^d \rangle_{q,\omega} =
\sum_{K.v \ne 0} 
\frac{\rho_0^2 K^2 \Delta_K \delta(\omega-K.v)}{\eta^2v^2(K_x+q_x)^2+c^2 q^4}
\end{equation}
The dynamic correlations are bounded due to the presence of
mass term $K_x v$ in the denominator. $u^d$
saturates at large distances, even in $d=2$ if $T=0$. $u^d$ is
smaller by a factor $a/R_x \ll 1$ compared to $u^s$ at
the length scales $R_x$. 
$u^d$ thus represents a small additional wiggling motion around the
ground state. The massive propagator in (\ref{dynamicequ}) 
is very different from a thermal one $1/q^2$.

Extension to realistic elastic energy, e.g. a triangular lattice in $d=2$
is straightforward. The static displacements, within the
random force model, are now
$u_{\alpha}(q) = F_y(q)( P^T_{\alpha y}(i \eta v q_x + c_{66} q^2)^{-1}
+ P^L_{\alpha y}(i \eta v q_x + c_{11} q^2 )^{-1})$
where $c_{11}$ and $c_{66}$ are (dispersionless) bulk
and shear moduli, respectively. Thus the mean square displacement $B(x,y)$ is
again given, for $y>y^*$, by (\ref{displ}) with $c$ replaced
by $c_{11}$. 
Note that only shear modes were considered in \cite{koshelev_dynamic},
an approximation which may hold for $y \ll y^*$
but misses the physics of the glass. Indeed only the {\it compression} modes
are responsible for the glass (and 
lead to unbounded displacements for $d>2$) since
both displacements and force have to be considered along y.
In $d=3$ tilt modes would also be relevant for flux lines, and
transverse shear modes for a solid. One finds 
$y^* = c_{11}/(\eta v) \sqrt{(c_{11}^2-c_{66}^2)/(c_{11}^2+c_{66}^2)}$.

The predictions of glassy structure, topological
order, channel motion and transverse critical force 
can be tested in experiments. For vortex lattices
$J_c^y$ can be measured in presence of a longitudinal
current. Magnetic noise experiments and NMR
probe $v t + u(x,t)$ and 
the phase at the washboard frequency 
should contain a static component with slow 
and anisotropic decay. 
Other experimental systems such as colloids, magnetic bubbles 
and double or triple incommensurate CDW should exhibit 
similar behavior. A transverse critical force
may explain recent Hall voltage experiments in 
2D Wigner crystals \cite{williams}.
The predictions of channel motion 
and transverse critical current can be directly tested 
in numerical simulations \cite{koshelev_simulations}.

(\ref{eqmotion2}) shows the 
importance of the relative orientation
of the lattice and the applied force.
It has been argued \cite{hauger_schmidt}
that to minimize power dissipation the
lattice aligns with the force. This process
may be slow and other orientations 
can be studied by applying a transverse field.
We expect commensuration effects with devil's
staircase type structure in the response to additional
force. At higher commensuration vectors
the channel structure may become unstable 
due to stronger effect of $u^d$. Conversely,
the larger the static part, the more 
stable the glass with fewer topological defects.
The size of the plastic flow regime should thus
depend on the lattice orientation. It is then possible 
that in $d=3$ for weak disorder the glass remains topologically 
ordered at all $v$ and that
the intermediate plastic regime disappears at low T.
One would thus go smoothly
from the moving to the static Bragg glass. At large 
$v$ channels are nearly straight and
out of equilibrium dislocations are thus suppressed. 
This allows to understand why in \cite{yaron_neutron} the plastic regime
disappears when $J$ is slowly decreased.

Previous descriptions
of moving systems, such as manifolds driven
in periodic \cite{spohn} or disordered potentials, focused
on the generation under RG of dissipative Kardar-Parisi-Zhang 
$(\nabla u)^2$ non linearities. They do occur,
due to lattice cutoff, in driven random sine-Gordon models
\cite{hwa_ledou,krug}. They are not important here because of the 
statistical symmetry 
$u_y \to - u_y$ in (\ref{staticequ})
and the fact that dynamic modes
(\ref{dynamicequ}) are massive. Thus, because
of periodicity this problem belongs to a new universality class. 
KPZ terms may play a role for incommensurate motion.

In conclusion, we studied a lattice moving
in a random potential. Static disorder dominates
motion along symmetry directions and the
moving system is a {\it glass} with a large amount
of topological order.
It is continuously related to the
static Bragg glass and although the decay of
translational order is slow it has
genuine glass properties different from a usual solid.
We predict experimental signatures
such as elastic channels and transverse critical current.

We thank Argonne National Laboratory 
for hospitality and NSF-Office of Science and Technology Center 
for support under contract DMR91-20000. We are
grateful to A. Koshelev and V. Vinokur for
stimulating and interesting discussions during the
making of this work, and a detailed explanation of
\cite{koshelev_dynamic}. PLD acknowledges discussions
with G. Crabtree and C. Bolle.


\begin{references}
\bibitem{perturb} the invalidity of perturbation theory for
transverse modes was noticed in \cite{hauger_schmidt}.
\end{references}
\end{document}